\documentclass[preprint,amsfonts,nofootinbib,aps]{revtex4}

\usepackage{graphicx}
\usepackage[T1]{fontenc}
\usepackage{calligra}

\begin{document}
\title{An axiomatic formulation of the
Montevideo interpretation of quantum mechanics}

\author{Rodolfo Gambini$^{1}$,
Luis Pedro Garc\'{\i}a--Pintos${^1}$,
and Jorge Pullin$^{2}$}
\affiliation {1. Instituto de F\'{\i}sica,
Facultad de Ciencias, Igu\'a 4225, esq. Mataojo, Montevideo, Uruguay. \\
2. Department of Physics and Astronomy, Louisiana State
University, Baton Rouge, LA 70803-4001}

\date{July 30th 2011}

\begin{abstract}
  We make a first attempt to axiomatically formulate the Montevideo
  interpretation of quantum mechanics. In this interpretation
  environmental decoherence is supplemented with loss of coherence due
  to the use of realistic clocks to measure time to solve the
  measurement problem. The resulting formulation is framed entirely in
  terms of quantum objects. Unlike in ordinary quantum mechanics,
  classical time only plays the role of an unobservable parameter. The
  formulation eliminates any privileged role of the measurement
  process giving an objective definition of when an event occurs in a
  system.
\end{abstract}
\maketitle

\section{Introduction}

The usual textbook presentation of the axiomatic formulation of
quantum mechanics includes two apparently unconnected problematic
issues.  The first one is the privileged role of the time variable
which is assumed to be a classical variable not represented by a
quantum operator. The second is the also privileged role of certain
processes called {\em measurements} where quantum states suffer abrupt
changes not described by a unitary evolution, and probabilities are
assigned to the values that one may obtain for a physical quantity.

The special role of measurement processes in quantum mechanics 
requires understanding what distinguishes such processes from the 
rest of the quantum evolution. This is called the {\em measurement
problem}, which many physicists have alluded to and that ultimately
refer to the uniqueness of macroscopic phenomena within a quantum
framework that only refers to potentialities. Ghirardi calls this 
the {\em problem of macro objectification}.

The orthodox response of the Copenhagen interpretation argues that the
objective of quantum mechanics is not to describe {\em what is} but
{\em what we observe}. The measuring devices are classical objects
through which we acquire knowledge of the quantum world. The
measurement therefore acquires an epistemological interpretation,
referring to processes in which observers acquire knowledge of
phenomena. The question about how does quantum mechanics account for
events observed in measurements and the multitude of events that
happen every moment in every place giving rise to the defined
perception of our experience is left out of the realm of the theory.
Those processes belong to a world of objects that our knowledge cannot
have access to. As put by d'Espagnat \cite{despagnatphil}, ``the
(orthodox) quantum formalism is predictive rather than
descriptive... [but also] ...the formalism in question is not
predictive (probability-wise) of {\em events}. It is predictive
(probability-wise) of {\em observations}.'' For him the statements of
quantum mechanics are {\em weakly objective} since they refer to
certain human procedures ---for instance, of observation---. They are
objective because they are true for everyone, ``But their form (or
context) makes it impossible to take them as descriptions of how the
things actually are''. Such descriptions are the usual ones in the
realm of classical physics, whose statements can be considered as {\em
strongly objective} since one can consider that they inform us about
certain attributes of the objects it studies.

If the statements of quantum mechanics can only be weakly objective
one must abandon attempts to understand how the passage from quantum
potentialities to observed phenomena, from micro to macro, from
determinism to randomness, from quantum to classical, takes place.
The question of which systems should be treated as classical also
becomes not analyzable, an issue that acquires more relevance as
more and more macro systems that display quantum behaviors are
being constructed by experimentalists.

If one adopts a realist point of view, that is, if one assumes the
existence of a reality independent of observers, the orthodox
description of quantum mechanics is incomplete since it does not tell
us which events may occur nor when may they occur. In our view this is a 
rather extreme point of view that should be reserved only to the case
in which one has exhausted all other possibilities for analyzing 
physically the problem of the production of events. There has been
a recent renewed interest among specialists in foundations of quantum
mechanics in understanding how an objective description at a macroscopic
level compatible with quantum mechanics arises. Several avenues
have been proposed to address such a question (for a comprehensive
review see \cite{ghirardireview}).

On the other hand the fact that time is treated unlike any other
variable in quantum mechanics has received much less attention. The usual
point of view is that to associate time with a quantum variable is
impossible. This is due to the well known Pauli observation that an
observable associated with time would be canonically conjugate to the
Hamiltonian and it is impossible to have a bounded below operator like
the Hamiltonian canonically conjugate to a self adjoint operator. Even
if one admits Leibniz' point of view that time is a relational notion
and therefore in modern terms described by clocks subject to the laws
of quantum mechanics, it is usually thought that this would only
complicate the description. The absolute Newtonian view imposed itself
not because it was the philosophically correct one but because it was
the simplest and yielded highly accurate predictions. A relational
treatment is only adopted if its use is inescapable, like in
situations where there obviously is no external parameter. An example
of this could be quantum cosmology where there are no external clocks,
nor external apparata to make measurements, nor an external
observer. As Smolin \cite{smolin} put it ``Can a sensible dynamical
theory [of quantum cosmology] be formulated that does not depend on an
absolute background space or time? Can quantum mechanics be understood
in a way that does not require the existence of a classical Observer
outside the system'?''  Up to now there have not been formulations of
theories of physics that are completely relational without unobservable
external elements.

The {\em Montevideo interpretation} \cite{montevideo} of quantum
mechanics shows that a relational treatment with quantum clocks allows
to solve the measurement problem, therefore providing a solution to
both the problems we mentioned above. In this paper we present an
axiomatic formulation of the Montevideo interpretation of quantum
mechanics where the evolution is described in terms of real clocks.
The formulation does not require the treatment of any observable as
classical or external.  In the axiomatic formulation we establish
precisely when and where events occur and what is their nature. Since
the formulation arises from an analysis of the problem of time in
quantum gravity \cite{time}, the proposed description ---although
presented here in the non-relativistic case only--- is formulated in a
language that is ready to treat generally covariant theories like
general relativity. It can be said that it is a quantum mechanics
formulated with an eye towards a quantum theory of gravity.


The axiomatic formulation has several goals: a) to give a rigorous
definition of what a {\em real clock} is; b) to list explicitly the
hypotheses of the Montevideo interpretation and to show its internal
consistency and c) to make explicit the mechanisms for macro
objectification and outline a realistic ontology based on this
interpretation. The resulting description will be strongly objective
in the sense indicated above without ever referring to observers
or measurements.  It does not attempt to substitute the usual
axiomatics in most practical applications, where the use of ideal
clocks gives a very precise description. An axiomatic relational 
formulation necessarily requires systems with enough degrees of
freedom to include the micro-systems\footnote{The typical 
systems with few degrees of  freedom one usually studies in quantum mechanics} one studies, the clocks, measuring
devices and the environment that is involved in the measurement
process.

\section{Axioms that are shared with ordinary quantum mechanics}

\subsection*{Axiom 1: States}

{\em The state of a complete physical system (including clocks, and if
  present, measuring devices and environment)
} {\calligra S} 
{\em is described 
by positive definite self-adjoint operators $\rho$ in a Hilbert space 
{\cal H}}

We adopt the idea that a state is well defined when it allows to
assign probabilities to any property associated with a physical
quantity. Examples of states are projectors on one-dimensional vector
subspaces, in which case the information contained in $\rho$ is
equivalent to that of a vector in the Hilbert space. The components of
the operator $\rho$ in a basis are usually referred to as the elements
of the {\em density matrix}.  The reason we are working with density
matrices is that as we will see, when one works with real clocks there
is loss of quantum coherence and this is more naturally discussed in
terms of density matrices.

The axiomatic formulation we are presenting makes reference to a set
of primitive concepts like {\em system, state, events and the
  properties that constitute them, and physical quantities}, each of
them associated with certain mathematical objects of the formalism of
ordinary quantum mechanics. All these are defined implicitly in the
axioms just like in ordinary axiomatic quantum mechanics one defines
system, state, measurement and physical quantities. The first axiom
associates certain operators to the states and a Hilbert space to the
systems.

\subsection*{Axiom 2: Physical quantities}

{\em Any physical quantity {\calligra A} of {\calligra S} is described by a
self-adjoint operator $\hat{A}$ that acts in {\cal H}. We will call
such operators observables}

In most situations, as we will see later, quantities of interest are 
associated to subsystems of {\calligra S}.

\subsection*{Axiom 3: Properties}

{\em The only possible values of a physical quantity {\calligra A} are 
the eigenvalues of the corresponding operator $\hat{A}$.}

 A physical quantity only takes values when an event occurs. If
 {\calligra A}\,\, has a value $A$ we will say that the event has a
 property {\calligra P}${}_A$ to which we will associate a projector
 $\hat{P}_A$ on the eigenspace associated with the corresponding
 eigenvalue $A$.

The events that constitute the physical phenomena are the most
concrete thing that attains us directly and we cannot ignore. They are
what makes the world and what physics has to account for.  It is
natural that physics, which is an empirical science would take as
starting point the events, which are the data from our experience of
the world. The word phenomenon comes from the Greek and means
something sufficiently apparent to be perceived by our senses. Events
are elementary phenomena that we usually associate with a set of
properties characterized by the numerical values that certain
physical quantities take, and their associated projectors. An example of
event would be the formation of a dot of silver atoms on a
photographic plate of an electron detector or the appearance of
droplets in a cloud chamber. In spite of the persistent tendency to
think in terms of particles in physics, we only observe events. 
The trace of a particle in a bubble chamber is just a series of
correlated events. Physical properties characterize events. For
instance, if we are interested in the position of the dot of silver on
the photographic plate, the position will be the physical quantity and
the value that it takes in a given experiment will correspond to a 
property that constitutes the event. Notice that we are not assuming
that all events are perceived by our senses.

\subsection*{Axiom 4: Evolution in Newtonian time}

In non-relativistic theories there exists a Newtonian time for
which the principle of inertia holds. That is, for which free
classical particles have a uniform rectilinear motion. Newtonian
time imposes an absolute order of events and an absolute notion of
simultaneity. Such an absolute time is not an accessible physical
quantity. It can only be approximately monitored by physical clocks,
which are subject to quantum fluctuations. This next axiom will refer
to the particularly simple form of the evolution of operators in
Newtonian time, which we will represent by a c-number $t$. We are
here working in the Heisenberg picture in which operators evolve.

{\em The evolution in Newtonian time of a physical quantity
with an associated self-adjoint operator $\hat{A}$ is given by the
equation}

\begin{equation}
i\hbar \frac{d \hat{A}(t)}{dt} = \left[\hat{A}(t),\hat{H}(t)\right]
+i\hbar \frac{\partial \hat{A}(t)}{\partial t}.
\end{equation}

For instance, in ordinary particle mechanics where one has its classical
position and momentum given by $x$ and $p$, an observable associated
with the classical quantity $A(x,p,t)$ is quantized by replacing $x$
and $p$ with $\hat{x}$ and $\hat{p}$ and appropriately symmetrizing so
that the resulting operator is self-adjoint. The partial derivative
refers to the explicit dependence in the parameter $t$. In ordinary
quantum mechanics the Heisenberg and Schr\"odinger pictures are
equivalent and so they are here if one is referring to the evolution in
terms of the (unobservable) Newtonian time $t$. If one considers the evolution
as described by real clocks there are modifications, as we will
subsequently discuss.

\section{Relational Axioms}
\label{relationalaxioms}
The probability axiom and the reduction axiom radically change their
form in the Montevideo interpretation since they now include the
observed system and the clock that registers the event, both as
quantum mechanical systems. We will consider ``almost uncoupled''
clocks, that is, weakly interacting with other degrees of
freedom. In order to simplify calculations, we
will also assume this means the clock degrees on freedom are not
entangled with other degrees of freedom: the Hilbert space of the
clock will be in a tensor product with the rest of the system. We
therefore say that a system contains a decoupled clock when the
Hamiltonian may be written in the form,
\begin{equation}
\hat{H}=  \hat{H}_{\rm clock}  + \hat{H}_{\rm system} , \label{3.2}
\end{equation}
where $\hat{H}_{\rm clock}$ depends only on the coordinates and momentum of
the clock and $\hat{H}_{\rm system}$ is independent of the clock
variables. While this situation is, strictly speaking, unphysical,  it 
approximates systems which differ from (\ref{3.2}) only
by terms that may be treated adiabatically. In correspondence with
this we will assume that the quantum state of the complete system is a
tensor product of a state for the clock and a state for the system
under study, i.e. $\rho=\rho_{\rm cl}\otimes \rho_{\rm sys}$ as stated
above.

A (linear) clock is a
dynamical system which passes through a succession of states at
constant time intervals. It can measure the duration of a physical
process and provides a quantitative description of the evolution.
Clocks have been introduced and analyzed by several authors
\cite{salecker,peres,buttiker,bonifacio}.  A recent review of the role
of time in quantum mechanics appears in \cite{hilgevoord}. These
authors have shown that dynamical position and time variables
---associated to rods and clocks--- are essentially of the same quantum
nature and that there is nothing in the formalism of quantum mechanics
that forces us to treat position and time differently.

Let $\hat{T}(t)$ be a self-adjoint operator (observable) in the
Hilbert space {\cal H} that describes the physical quantity chosen to
measure time by a clock ruled by quantum mechanics and $\hat{Q}^i(t)$
and $\hat{P}^i(t)$ observables associated to set of quantities {\calligra Q}
and {\calligra P} \, that commute with $\hat{T}(t)$ and whose
values one wishes to assign probabilities to. We assume all variables
have continuous spectrum, because clocks normally do, results are
easily reworked for variables having discrete spectrum.  Let
$\hat{P}_{Q^i_0}(t)$ be the projector on the eigenspace of $\hat{Q}^i$ with
eigenvalues in the interval of a given width $2\Delta^i$ centered in
$Q^i_0$, that is, $[Q^i_0-\Delta^i,Q^i_0+\Delta^i]$ and analogously
the clock variable $\hat{T}$ with its projector $\hat{P}_{T_0}(t)$. In terms of
these quantities the probability postulate states that:

\subsection*{Axiom 5: probabilities}

{\em The probability that the quantity {\calligra Q}\,\,\,\,$^i$ of a
physical system in a state $\rho$ take a value in a prescribed range
of values when the clock in such state takes a value in the interval
$[T_0-\Delta^C,T_0+\Delta^C]$ is given by,}
\begin{equation}
{\cal P}_C\left(Q^i\in \left[Q^i_0-\Delta^i,Q_0^i+\Delta^i\right] \vert
T\in \left[T_0-\Delta^C,T_0+\Delta^C\right] \right) =
\frac{\int_{0}^\tau dt 
{\rm Tr}\left(\hat{P}_{Q^i_0}(t) \hat{P}_{T_0}(t) \rho \hat{P}_{T_0}(t)\right)}
{\int_{0}^\tau dt 
{\rm Tr}\left(\hat{P}_{T_0}(t) \rho\right)},\label{probability}
\end{equation}
where $\hat{P}_{Q_0}(t)$ and $\hat{P}_{T_0}(t)$ are the projectors
associated to properties $Q$ and $T$ taking the eigenvalues $Q_0$ and 
$T_0$. 

These conditional probabilities are positive and add to one. They
refer to the probability of occurrence of events with properties
associated with the eigenvalues of the operators involved. Which
specific events and when do they occur are issues not determined by
this axiom. Notice that a similar construction can be carried out for
the {\calligra P}\,\,\,\,$^i$ quantities, we wrote the expression for
the {\calligra Q}\,\,\,\,$^i$ for concreteness only. The only
condition is that the quantities must have vanishing Poisson bracket
with $T(t)$.

Note that we are integrating in the Newtonian time $t$ which is taken
to be unobservable. The integration interval goes from $t=0$, instant
in which the observable clock $T$ is started, to $\tau$, the maximum
Newtonian time for which the clock $T$ operates with a given
precision.  The quantity $\tau$ makes reference to the interval in
which the clock is operational, and therefore in that sense the left
hand side of (3) depends on $\tau$. No physical clock can operate
indefinitely.  The quality of the clock depends on its initial state
when it is started, its dynamics, the admissible error $\Delta^C$ and
the total time the clock is used $\tau$.  The probabilities assigned
in axiom 5 are therefore clock-dependent in various ways and we denote
that with the subindex $C$.

If one wishes to perform subsequent
measurements care should be taken to choose the interval $\Delta^C$
large enough such that the measurement of the clock variable does not
affect too much the accuracy of it.  Later on, we will obtain
ontological realistic conclusions from this axiom in spite of its
clock dependence, since there exist physical bounds on the accuracy of
clocks \cite{ng} independent of any observer. The notion of
undecidability we will introduce later will refer to those bounds and
therefore will be clock independent.

As we argued in
\cite{time}, ``
It is worthwhile expanding on the meaning of the probabilities
(\ref{probability}) since there has been some confusion in the
literature \cite{hu}. Thinking in terms of ordinary quantum mechanics
one may interpret that the numerator of (\ref{probability}) is the sum
of joint probabilities of $Q$ and $T$ for all values of $t$. This
would be incorrect since the events in different $t$'s are not
mutually exclusive. The probability (\ref{probability}) corresponds to
a physically measurable quantity, and  such quantity is actually
the only thing one can expect to measure in systems where one does not
have direct access to the (unobservable)  time $t$. The experimental setup
we have in mind is to consider an ensemble of non-interacting systems
with two quantum variables each to be measured, $Q$ and $T$. Each
system is equipped with a recording device that takes a single
snapshot of $Q$ and $T$ at a random unknown value of the (unobservable)
time $t$. One takes a large number of such systems, launches them all
in the same quantum state, ``waits for a long time'', and concludes
the experiment. The recordings taken by the devices are then collected
and analyzed all together. One computes how many times $n(T_j,Q_j)$
each reading with a given value $T=T_j,Q=Q_j$ occurs (to simplify
things, for the moment let us assume $T,Q$ have discrete spectra; for
continuous spectra one would have to consider values in a small finite
interval of the value of interest). If one takes each of those values
$n(T_j,Q_j)$ and divides them by the number of systems in the
ensemble, one obtains, in the limit of infinite systems, a joint
probability $P(Q_j,T_j)$ that is proportional to the numerator of 
the above expression.'' The denominator is obtained by counting 
$n(T_j)$ ignoring the values of $Q$. Notice that this implies a change
in the probability axiom with respect to ordinary quantum mechanics.
This is what is made explicit in axiom 5.

The previous expression can be straightforwardly extended to the case
in which one or both observables involved have discrete spectrum.
Since the spectrum may be time dependent it is also convenient to talk
about quantities taking values in finite intervals in the discrete
case as well.

Although we spelled out the axiom explicitly for the measurement of a
single quantum observable $\hat{Q}^i$ it is immediately generalizable to the
measurement of several commuting operators (functions of the $\hat{Q}^i$'s
and $\hat{P}^i$'s). The next axiom allows to assign probabilities to
histories of events that occur at different instants of time.

\subsection*{Axiom 6: State reduction}

{\em When a set of physical quantities (that include the clock) with
commuting self-adjoint operators $\hat{A}_1\ldots \hat{A}_n$ take
values ${A}^1\ldots {A}^n$ in the intervals
$[A^1_0-\Delta^1,A^1_0+\Delta^1]\ldots
[A^n_0-\Delta^n,A^n_0+\Delta^n]$ the state of the system can be
represented by the normalized quasi-projection of the state $\rho$
associated with the values of the quantities in question,}
\begin{equation}\rho_{\rm red}=
\frac{\int_0^\tau dt \hat{P}_{A^1_0}(t)\ldots \hat{P}_{A^n_0}(t) \rho 
\hat{P}_{A^n_0}(t)\ldots \hat{P}_{A^1_0}(t) }
{{\rm Tr}\left(\int_0^\tau dt\hat{P}_{A^1_0}(t)\ldots \hat{P}_{A^n_0}(t) \rho 
\hat{P}_{A^n_0}(t)\ldots \hat{P}_{A^1_0}(t) \right)}.\label{quasiprojector}
\end{equation}

This is a quasi-projection\footnote{A quasi projector is a self
  adjoint operator having only discrete eigenvalues lying in the
  interval $[0,1]$. The idea is that it has many eigenvalues near $1$,
  relatively few between $0$ and $1$ and many close to zero. More
  precisely a quasi projector of rank $N$ and order $\eta$ satisfies
  ${\rm Tr}(F)=N$ and ${\rm Tr}(F-F^2)=N O(\eta)$ with $\eta<<1$}
 (as defined by Omn\'es \cite{omnes}) since
it is not an exact projector. If one were able to have an uncoupled clock,
that is, if the total Hilbert space could be written as the tensor
product of the Hilbert space of the clock times the Hilbert space of
the rest of the system, then the probability density given by,
\begin{equation}
{\cal P}_t(T)=
\frac{{\rm Tr}\vert_{\rm cl}\left(\hat{P}_T(t) \rho_{\rm cl}\right)}
{\int_0^\tau  dt {\rm Tr}\vert_{\rm cl}(\hat{P}_T(t) \rho_{\rm cl})}
\label{probabilityt}
\end{equation}
would be a Dirac delta ${\cal P}_t(T)= \delta(t-T)$ and
(\ref{quasiprojector}) would behave as an exact projector when the
reduction postulate is used to assign probabilities to histories
\cite{obregon}.  ${\cal P}_t(T)$ is the probability density that the
unobservable time takes the value $t$ when the physical clock reads $T$, and
is not a directly observable quantity in our framework (since $t$ is
not observable) but a mathematical object that appears in intermediate
calculations.

This axiom only has epistemological character, it does not say that
the state actually undergoes the above mentioned reduction process. In the
present theory if the state does or does not undergo reduction is an
undecidable proposition, as we will discuss in the next section. 

Using the same construction as in ordinary quantum mechanics of combining
the reduction and the probability axioms one can assign probabilities
to histories of events. In \cite{time} we showed in model systems that
the resulting probabilities of histories can be used to construct the
ordinary particle propagator to leading order in the inaccuracy of the
clock. This is true even for generally covariant systems like general
relativity, resolving a longstanding issue in the definition of a
notion of time for such systems.

Introducing a reduction postulate superficially seems to leave the
measurement problem intact. Up to this point, the relational
description of evolution presented does not provide information about when
events occur. Notice that one cannot simply say that events happen
randomly since generically they lead to a $\rho_{\rm red}$ that is
physically distinguishable from $\rho$ and that would completely
destroy the predictive power of quantum mechanics. As Bell noted, this
would be the situation in ordinary quantum mechanics if we adopted the
language of events instead of that of measurements.  The main
difference in the current axiomatic system, as we will show, is that
it allows situations where the events can occur and gives a physical
criterion to establish when they occur. The next and final axiom will
be crucial for this issue.

\section{Axiom 7: Fundamental limitations in measurements and the
  ontological axiom}

\subsection{Loss of unitarity due to the use of real clocks}

In preparation to formulate the seventh axiom, 
we would like now to address a new phenomenon: the loss of unitarity
of quantum mechanics described with real clocks. Let us reconsider
the conditional probability (\ref{probability}),
\begin{equation}
{\cal P}\left(Q^i\in \left[Q^i_0-\Delta^i,Q^i_0+\Delta^i\right] \vert
T\in \left[T_0-\Delta^C,T_0+\Delta^C\right] \right) =
\frac{\int_{0}^\tau dt 
{\rm Tr}\left(\hat{P}_{Q^i_0}(t) \hat{P}_{T_0}(t) \rho \hat{P}_{T_0}(t)\right)}
{\int_{0}^\tau dt 
{\rm Tr}\left(\hat{P}_{T_0}(t) \rho\right)},
\end{equation}
and make some reasonable assumptions about the clock and the system as
we discussed in section \ref{relationalaxioms}.  Going to the
Schr\"odinger picture we define a new density matrix for the system
excluding the clock labeled by the
physical time $T$ instead of the unobservable Newtonian time $t$,
\begin{equation}
\rho_{\rm sys}(T)\equiv\int_0^\tau dt {\cal P}_t(T) \rho_{\rm sys}(t) \label{definerho}
\end{equation}
where ${\cal P}_t(T)$ was defined in (\ref{probabilityt}).
In terms of these density matrices the conditional probability can
be rewritten as,
\begin{equation}
{\cal P}\left(Q^i\in \left[Q^i_0-\Delta^i,Q^i+\Delta^i\right] \vert
T\in \left[T_0-\Delta^C,T+\Delta^C\right] \right) =
\frac{{\rm Tr}\vert_{\rm sys}\left(\hat{P}^S_{Q^i_0} \rho_{\rm sys}(T)\right)}
{{\rm Tr}\vert_{\rm sys} \left(\rho_{\rm sys}(T)\right)},
\end{equation}
where $\hat{P}^S_{Q^i_0}$ is the projector in the Schr\"odinger
picture.  We therefore see that we have recovered the ordinary
definition of probability of measuring $Q^i$ at time $T$ in usual
quantum mechanics. This shows the usefulness of the definition
(\ref{definerho}). Within such definition one can immediately see the
root of the loss of unitarity when one uses real clocks to describe
quantum mechanics. The density matrix in the right hand side of
(\ref{definerho}) evolves unitarily in the unobservable time $t$. However,
due to the presence of the probability ${\cal P}_t(T)$ the left hand
side does not evolve unitarily.  If one starts with a pure state, in
the right hand side it will remain pure, but in the left hand side
after some time has evolved one will end up with a mixture of
pure states due to the integral. Only if the probability ${\cal
P}_t(T)$ were a Dirac delta one would have a unitary evolution. That
would mean that one has a clock that correlates perfectly with $t$,
which is not possible with a real clock.

We therefore see that the result of Axiom 5 is to have a theory 
that looks like ordinary quantum mechanics but in terms of the 
physical time $T$. The only difference is that the evolution in terms
of the physical time is only approximately unitary. If one assumes that
the clock is very good the probability ${\cal P}_t(T)$ will be a Dirac
delta with small corrections,
\begin{equation}
{\cal P}_t(T) = \delta(T-t) + a(T) \delta'(T-t) + b(T) \delta''(T-t)+\ldots
\end{equation}
and one can show that in such a case the density matrix evolves
according to the equation,
\begin{equation}
i\hbar \frac{\partial \rho}{\partial T} = \left[\hat{H},\rho\right]
+\frac{\partial b(T)}{\partial T}
\left[\hat{H},\left[\hat{H},\rho\right]\right]
\end{equation}
so we see that to leading order we get the ordinary Schr\"odinger
evolution and the first corrective term has to do with the rate of
spread of the width of the probability ${\cal P}_t(T)$ plus higher
order corrections.  Another way of putting it is that it is determined
by how inaccurate the physical clock becomes over time. The effect can
therefore be reduced by choosing clocks that remain as accurate as
possible over time. However, there exist fundamental physical
limitations to how accurate one can keep a clock over time. There
are several arguments in the literature \cite{ng}  that suggest that the best
accuracy one can achieve with a clock is given by $\delta T \sim T^a
T_{\rm Planck}^{1-a}$ and $T_{\rm Planck}=10^{-44}s$ is Planck's
time. The estimates for $a$ vary but several authors claim it is
$1/3$. From the point of view of the purposes of this paper, it
suffices to say that $\delta T$ is a growing function of $T$. Then
unitarity is inevitably lost.

There have been attempts to bypass these limitations and construct
clocks whose inaccuracy does not grow with time.  Those attempts, as
for instance the Larmor clock \cite{peresbook} produced by using a
finite-dimensional quantum dial, are not physically
implementable. This particular one involves an infinite mass rigid
rotator.  All physically implementable linear clocks proposed up to
present have uncertainties in the measurement of time that grows with
time.

The fundamental bounds on the accuracy of physical clocks follow
from a joint consideration of quantum mechanics and general
relativity. If one were able to start from an axiomatic formulation of
quantum gravity they would not imply an additional hypothesis. However
as these considerations play a crucial role in the fundamental loss of
coherence that leads to the production of events, this assumption
should be stated explicitly as an

{\bf Auxiliary axiom:} {\em There is a fundamental uncertainty in the
measurements of time that grows with a positive fractional power a of
the time interval $\delta T=T^a T_{\rm Planck}^{1-a}$.}

The loss of coherence due to imperfect clocks makes the off-diagonal
elements of the density matrix of a quantum system in the energy
eigen-basis  decrease exponentially. For $a=1/3$, the exponent for
the $mn$-th matrix element is given by $\omega_{mn}^2 T_{\rm
Planck}^{4/3} T^{2/3}$, where $\omega_{mn}=E_{mn}/\hbar$ is the
difference of energy between levels $m$ and $n$ divided by $\hbar$
(the Bohr frequency between $n$ and $m$). One could see this effect in
the lab in reasonable times (hours) only if one were handling
``macroscopic'' quantum states corresponding to about $10^{13}$ atoms
in coherence.  The direct observation of this effect is therefore
beyond our current experimental capabilities. However, it has profound
implications at a foundational level, as this new formulation of
quantum mechanics we are presenting attests to. It should be noted
that what is not currently observable experimentally is the
fundamental limit to the accuracy of clocks. The effect associated
with the loss of coherence in realistic clocks can be made arbitrarily
large by choosing inaccurate clocks and has been observed
experimentally in ion traps \cite{bonifacio}.

\subsection{Undecidability}

The loss of unitarity due to the inaccuracies of real clocks has
implications for the usual explanation of the measurement process
through environmental decoherence. The results of such program can be
summarized as follows: consider a system {\calligra S}\,\, interacting
with an environment {\calligra E}\, with a total Hamiltonian
$\hat{H}=\hat{H}_{SA}+\hat{H}_E+\hat{H}_{int}$ with $\hat{H}_{SA}$ the Hamiltonian of the micro-system, which may include a measuring apparatus, $\hat{H}_E$
that of the environment  and
$\hat{H}_{\rm int}$ the interaction Hamiltonian between the system and the
environment. The effect of such interaction is an attenuation of the
interference terms in the reduced density matrix of the system
{\calligra S}\,\,, obtained by partially tracing over the degrees of
freedom of the environment. This effect happens in the so-called
``pointer basis'', determined by the Hamiltonian, as has been
discussed in some detail in \cite{pazzurek}.  The interpretation of
this attenuation is as follows: when one carries out local measurements on
the system {\calligra S}\,\, it will behave classically, any expectation
value will be equal to the case in which the system has suffered a
state reduction, and we cannot see the typical interference terms of
quantum superpositions. Since interactions with the environment are
almost inevitable, this is the reason why the world we experience everyday 
behaves classically and quantum behavior can only be directly seen in very
controlled circumstances in the lab. This is therefore portrayed as a
solution to the measurement problem.

There exist three limitations that have been pointed out in the
literature that may preclude some people from accepting that
environmental decoherence is a solution to the measurement
problem. The first two limitations are related to the fact that the
evolution for the total system {\calligra S}\,\, plus {\calligra E}\,
is unitary. The first limitation is the possibility of {\em
  revivals}. That is, for a closed total system one could wait for a
long time and see the quantum coherence in the system {\calligra S}\,
plus the measuring apparatus reappear. The use of real clocks prevents
this from happening, since waiting for very long actually increases
the loss of coherence due to the clocks. The second limitation,
suggested in \cite{despagnat}, argues that one could perhaps construct
global observables that depend on variables in the system and the
environment that would suffer different changes in their expectation
values if a reduction takes place or not. A detailed analysis
\cite{gapipu} in model systems shows that one is prevented from
measuring such observables when one takes into account the loss of
coherence due to real clocks. The third limitation to viewing the use
of environmental decoherence as a solution to the measurement problem
is that ``nothing happens'', that is, there is no criterion given for
telling when an event (or a measurement) takes place. The fact that
the reduced matrix of the open subsystem composed by the micro-system
and the measurement device takes a diagonal form does not change the
interpretation of the state as a superposition of options. This is
what Bell called ``the and/or problem'' alluding to the lack of
justification for assuming that a transition from superposed options
to alternative options takes place. We will resolve this in our
approach by providing a criterion for when an event takes place.

Returning to the first objection, one may ask 
how many degrees of freedom one needs to consider for the
exponential decrease to kill the possibility of revivals? A criterion
would be that the magnitude of the off diagonal term in the revivals
be smaller than the magnitude of the off diagonal terms in the
intermediate region between revivals. If that were the case the
revival would be less than the ``background noise'' in regions where
there is no revival. The magnitude of the interference terms in the
density matrix were studied by Zurek \cite{zurek} in a simple model
with two levels where the environment is characterized as $N$
particles, and goes as $\rho_{+-}
\sim 1/2^{N/2}$ with $N$ the number of particles. 
The time for revivals to occur goes as $T\sim N!$. This implies, at
least in this particular example, that if one has more than hundreds
of particles in the environment the loss of coherence will make
the observation of revivals impossible. In realistic environments the
number of degrees of freedom is of course vastly higher.

As was discussed in \cite{gapipu}, it is worthwhile emphasizing the
robustness of this result in practical terms. One could, for instance,
question how reliable the fundamental limits for the inaccuracy of
clocks we are considering are. Some authors have characterized the
fundamental limit as too optimistically large, arguing that the real
fundamental limit should not  be larger than  Planck time
itself. In view of this it is interesting to notice that if one posits
a much more conservative estimate of the error of a clock, for
instance $\delta T \sim T^\epsilon T_{\rm Planck}^{1-\epsilon}$, for
any small value of $\epsilon$ the only modification would be to change
the number of particles $N_0\sim 100$ to at least $N\sim
N_0/(3\epsilon)$. So the only real requirement is that the inaccuracy
of the clock increases with the time measured, a very reasonable
characteristic for any realistic clock.

Using a real clock introduces a fundamental difference. Whereas in the
usual formalism the state of the system plus apparatus plus
environment will evolve unitarily, here it will lose coherence without
the possibility of recovering it in another part of the system.  This
brings us to the idea of {\em undecidability}. If a system suffers an
interaction such that one cannot distinguish by any means if a unitary
evolution or a reduction took place we will claim that an {\em event
  took place}. This provides a criterion for the production of events,
as we had anticipated.  We will provide a detailed form of the
criterion later on. Notice that for a quantum micro-system in
isolation, events would not occur. However for a quantum system
interacting with an environment, events will be plentiful. The same
goes for a system being measured by a macroscopic measuring device.
It should be emphasized that the notion of undecidability is
independent of a particular clock, since it is based on the best
possible clock. Precisely, the situation becomes undecidable when the
distinction is impossible for any physical clock. This is the reason
why the fundamental limitations for the measurement of time intervals
mentioned above become important.

\subsection{Axiom 7: The ontological axiom}

The analysis of the previous section shows that contrary to what
happens in quantum mechanics with an ideal clock, in the relational
picture the possibility to determine (not just in practice but in
principle) if a system has suffered a state reduction or evolved
unitarily decreases exponentially with the number of degrees of
freedom of the system. That is, it requires to consider ensembles with a number
of identical macroscopic systems exponential in the number of degrees
of freedom of the total system including environment and measuring
apparatus. One cannot therefore argue ---as is done in the case of
ordinary environmental decoherence--- that the problem moves on to the
complete system that retains the complete initial quantum
information. The existence of this phenomenon in systems that interact
with an environment implies, as follows from the above analysis, that
in processes where there does not exist an unlimited capability of
preparing the initial state of the system it will be undecidable if
there irrespective of a reduction taking place (or not). By
undecidable we mean that the expectation values of any observable of
{\calligra S}\,\, will be identical in both cases.

This leads to the following ontological axiom that gives sufficient
physical conditions for the production of an event. We lay it out for
variables with continuous spectrum but it is readily generalizable to
other cases. The axiom reads:

{\em Consider a closed system {\calligra S} with its associated Hilbert
  space {\cal H} and a physical quantity {\calligra A} represented by
  an observable $\hat{A}$ in {\cal H} with a decomposition of the
  identity allowing to write $\hat{A}(t)=\sum_n a_n
  \hat{P}_{a_n}(t)$. We will say that 
  an event occurs when it becomes impossible to distinguish (in terms
  of the expectation values of any observable quantity), in a certain
  instant in which the clock reads in an interval $2\Delta^C$ centered
  in $T_0$, between the initial state\footnote{Notice that we are in
    the Heisenberg representation. In the Schr\"odinger representation
    it would be the density matrix at time $t$ modified.} of
  {\calligra S}\,\, modified by the clock reading},
\begin{equation}
  \rho_{\rm mod}=\frac{\int_0^\tau dt \hat{P}_{T_0}(t)\rho
    \hat{P}_{T_0}(t)}
{\int_0^\tau {\rm Tr}\left(\hat{P}_{T_0}(t) \rho\right)}
\end{equation}
{\em  and},
\begin{equation}
\rho_{\rm e} = \frac{\int_0^\tau dt \sum_n \hat{P}_{a_n}(t)
  \hat{P}_{T_0}(t) \rho \hat{P}_{T_0}(t) \hat{P}_{a_n}(t)}
{\int_0^\tau  dt {\rm Tr}\left(\hat{P}_{T_0}(t)\rho\right)}.
\end{equation}
{\em The event associated with the physical quantity {\calligra A}\,\,
taking the value $a_n$ occurs with a probability given by axiom 5.
Such event will have a property associated with the projector
$\hat{P}_{a_n}(t)$ with relative probability ${\cal P}_t(T_0)$.} Notice
that $\rho_{\rm e}$ is the density matrix that one would have after a
traditional wavefunction collapse and that $\rho_{\rm mod}$ and
$\rho_{\rm e}$ are states in the Hilbert space of the system plus
environment.

We are assuming that we have a good clock that works with a certain
degree of accuracy for a period of Newtonian time $\tau\gg T_0$. With
this hypothesis the above construction is independent of $\tau$.  It is not
possible to assign a single property to the observation of $a_n$ since
the clock does not allow to identify a single projector due to the
ambiguity in the value of the unobservable time $t$ in which the event
occurs. In realistic situations, with good clocks, such ambiguity does
not have practical consequences since the variation of
$\hat{P}_{a_n}(t)$ in the interval
$\left[T_0-\Delta^C,T_0+\Delta^C\right] $ will be negligible.

As explained above, an event occurs when one cannot 
distinguish the physical 
predictions\footnote{To be mathematically precise, given 
the states $\rho_{\rm mod}$ and $\rho_{\rm e}$ and any property of 
{\calligra S}\,\, given by a projector $\hat{P}$ one has that
$\vert{\rm Tr}\left(P\left(\rho_{\rm mod}-\rho_{\rm
      e}\right)\right)\vert
<\epsilon$ with
$\epsilon=\exp(-\alpha N)$. $\alpha$ is a positive constant and 
$N$ the number of particles in the system (for an example see \cite{dice2010}). Notice that the term on the
left of the inequality is clock dependent. We request that the
inequality be satisfied for the best possible clock.}
of the modified density matrix $\rho_{\rm mod}$ and
the ones given by $\rho_{\rm e}$. 
This situation arises typically in systems
that interact with an environment with a large number of degrees of
freedom.  When this happens the physical quantity characterized by $\hat{A}$
will take a definite value. As we have emphasized, {\calligra S}\,\,
includes the micro-system and the environment with which it has
interacted.

In general many observables will satisfy the above condition, and
therefore many properties of the system will actualize.  To illustrate
this point we will consider a simplified situation.  Let us assume
that after the process of decoherence has been completed, the only
Hamiltonian present is that of the clock and that the system does not
evolve, so that we have time independent projectors,
\begin{equation}
\rho_e =  \sum_n \hat{P}_{a_n} \left( \frac{\int_0^\tau dt \hat{P}_{T_0}(t) \rho
    \hat{P}_{T_0}(t)}
{\int_0^\tau  dt {\rm Tr}\left(\hat{P}_{T_0}(t)\rho\right)}
 \right) \hat{P}_{a_n} \equiv \sum_n \hat{P}_{a_n} {\rho}(T_0) \hat{P}_{a_n}, \label{matrix}
\end{equation}
and it should be noted that $\rho(T_0)$ is the density matrix of the
{\em complete system}, in the Schr\"odinger picture labeled by the real
clock time $T_0$.  We will show that the condition for an observable
$B$ to also actualize is that its projectors' eigen-spaces include the
eigen-spaces of $A's$ projectors. That is,
\begin{equation}
\hat{P}_{b_n} \hat{P}_{a_n}^{} \vert \psi \rangle = \hat{P}_{a_n}^{} 
\vert \psi \rangle, \label{cond1}
\end{equation}
and
\begin{equation}
\hat{P}_{b_m} \hat{P}_{a_n}^{} \vert \psi \rangle = 0; \ \ m \ne n. \label{cond2}
\end{equation}
When the above conditions are satisfied we will say
that the projector $\hat{P}_{a_n}^{}$ includes $\hat{P}_{b_n}$, and that the
property corresponding to the first includes the second, {\calligra
  P}${}_{b_n} \subset$ {\calligra P}$\,\,{}_{a_n}^{}$ .

Let us assume that we have undecidability,
\begin{equation}
\rho_e = \sum_{n} \hat{P}_{a_n}^{} {\rho}(T_0) \hat{P}_{a_n}^{}, 
\end{equation}
then we will see that for observable $B$ the undecidability condition
is also satisfied.

Using the closure relationship we have that,
\begin{equation}
\sum_n \hat{P}_{b_n} {\rho}(T_0) \hat{P}_{b_n} = \sum_n \hat{P}_{b_n} \left( \sum_{k} \hat{P}_{a_k}^{} \right) {\rho}(T_0) \left( \sum_{l} \hat{P}_{a_l}^{} \right) \hat{P}_{b_n},
\end{equation}
and together with (\ref{cond2}) imply,
\begin{equation}
\sum_n \hat{P}_{b_n} {\rho}(T_0) \hat{P}_{b_n} = \sum_n  \hat{P}_{b_n} 
\hat{P}_{a_n}^{} {\rho}(T_0) \hat{P}_{a_n}^{} \hat{P}_{b_n}.
\end{equation}
Now using (\ref{cond1}) we have that,
\begin{equation}
\sum_n \hat{P}_{b_n} {\rho}(T_0) \hat{P}_{b_n} = \sum_n \hat{P}_{a_n}^{} 
{\rho}(T_0) \hat{P}_{a_n}^{} = \rho_e,
\end{equation}
and therefore $B$ is also undecidable.

We will call ``essential property'' the one that includes all
properties that actualize, that is, all properties whose projectors
satisfy the undecidability condition. This ``essential property''
contains the information of every physical quantity that the system
acquires.

Let us see how this works more explicitly in a simple example. We will
consider a system composed of only three spins, and the clock. Let
us assume that the initial state for the spins is
\begin{eqnarray}
  \rho(0) =  \frac {\vert c_1 \vert ^2}{2} 
\left( \vert ++- \rangle + \vert +-+ \rangle \right) 
\left( \langle ++- \vert + \langle +-+ \vert \right) + 
\vert c_2 \vert ^2 \left( \vert -++ \rangle  \right) 
\left( \langle -++ \vert  \right) \nonumber\\ + \frac{c_1 c_2^*}
{\sqrt 2} \left( \vert ++- \rangle + \vert +-+ \rangle \right) 
\left( \langle -++ \vert  \right) + \frac{c_1^* c_2}{\sqrt 2} 
\left( \langle -++ \vert  \right) \left( \langle ++- \vert + 
\langle +-+ \vert \right).
\end{eqnarray}

Suppose that the evolution is such that an event occurs\footnote{for
  small systems like the one we are considering events will not occur
  in general, since there is no undecidability.}, with essential
properties characterized by,
\begin{equation}
\hat{P}_{a_1} = \left( \vert ++- \rangle + \vert +-+ \rangle \right) 
\left( \langle ++- \vert + \langle +-+ \vert \right),
\end{equation}
and
\begin{equation}
\hat{P}_{a_2} = \left( \vert -++ \rangle  \right) \left( \langle -++ \vert  \right).
\end{equation}

As we noticed before, if for instance the property given by $\hat{P}_{a_1}$ is
attained, it gives all the information about the physical quantities
the system has. We can now consider the compatible property associated
with the projector,
\begin{equation}
\hat{P}_{\rm up} = \vert + \rangle \langle + \vert \otimes I_2 \otimes I_3,
\end{equation}
which corresponds to ``spin 1 is up''. And we could also consider the
compatible property associated to
\begin{equation}
\hat{P}_{{\rm 2opposite3}} = I_1 \otimes \left( \vert +- \rangle + \vert -+ \rangle \right) \left( \langle +- \vert + \langle -+ \vert \right)
\end{equation}
which corresponds to ``spins 2 and 3 are opposite''. Both $\hat{P}_{\rm up}$
and $\hat{P}_{\rm 2opposite3}$ satisfy condition (\ref{cond1}), so these
properties will actualize.

The projectors compatible with the essential properties determine the
properties that can be associated to different subsystems. So, in the
case of the property corresponding to $\hat{P}_{a_1}$ being acquired by the
system, we can ask whether spin 1 is up or not, we can ask whether
spins 2 and 3 are opposite or not, but we cannot ask whether spin 2 is
up or not, because this last property is incompatible and is therefore
not acquired by the subsystem.

Usually the essential property acquired by the system is complicated
and not experimentally accessible, but we are
interested in properties acquired by the subsystems when events occur.

The ontological axiom completes the formulation of the Montevideo
interpretation of quantum mechanics. It eliminates the need to give
special treatment to measurements and observers and gives rise to an
objective description completely independent of cognizant beings.

The reader may question what is the situation in an actual measurement
in the lab. There we have the possibility of forcing the occurrence of
events by designing a measuring apparatus/environment combination that
interacts with the system under study in such a way that the pointer
basis corresponds to eigenstates of the observable one desires to
measure. The effects discussed above occur and an event takes
place. The measurements discussed in quantum mechanics textbooks
therefore reduce to finding the correct Hamiltonians so that the
properties that actualize their values correspond to the observables
that one wishes to measure in each case.

\section{The role of states: do they describe systems or ensembles?}

What happens with the states? As we observed, it is not empirically
decidable what happens with the states when an event occurs. Although
the interpretation is compatible with a state of the universe given
once and for all, for practical purposes we will not have predictive
power if we do not know all the actualizations of events prior to the
moment of interest.  Due to this it will be convenient (and possible)
from the epistemological point of view to postulate that a reduction
takes place after the observation of an event. As Omn\`es points out:
``reduction is not in itself a physical effect but a convenient way of
speaking'' \cite{omnes}. More precisely, in the construction presented
in this paper it is not physically decidable if the reduction of the
state takes place or not. This is precisely the condition, as established in 
axiom 7, for events to occur.

If it were the case that the state undergoes an effective reduction
process every time an event occurs, then the state can be 
associated at all times with an individual system and knowledge of 
the state represents the maximum information available to make
predictions about future behaviors. 

If one adopts the opposite point of view and assumes that the state
remains unchanged during the processes in which events occur, the
state ---which would be none other than the initial state of the universe---
would describe ensembles of systems in which in every member of the
ensemble events of different nature would occur. In this case in order
to have complete information about the future behavior of the universe
would require not only knowledge of the state but all the events that 
have occurred previously to the instant in which one wishes to have
the information. It is important to notice here that the proposed
formulation would be complete without axiom number 6. It only has
the purpose of resolving the ambiguity noted above in order to use
the information added by the occurrence of the event in future 
predictions. Axiom 6 is therefore, as we have mentioned,  
of epistemological character. It allows to actualize the information
available after each measurement.

We have limited ourselves to closed systems. The systems have to be
general enough to include the various subsystems involved in the
occurrence of the events of interest. Some subsystems are agents that
initiate the process, like the electron in the double-slit
experiment. Others are recipients of the action, like the photographic
plate in that experiment. The total systems will only allow a complete
description of {\em some} processes that lead to events in {\calligra
  S}\,\,. We are able to describe events in which the system
{\calligra S}\,\, contains as subsystems. the quantum micro-system,
the environment and perhaps a measuring device.
There might be situations in which subsystems of {\calligra
S}\,\, act or are acted upon by subsystems not included in {\calligra
S}\,\,.  Events and states have a primary ontological status whereas
the systems considered here have circumstantial character and are
considered as long as they support the events and states of interest.

\section{Conclusions}

We have presented an axiomatic formulation of the Montevideo
interpretation of quantum mechanics. In this interpretation
environmental decoherence is supplemented with a fundamental mechanism
of loss of coherence due to the inaccuracy in tracking time that real
clocks introduce to produce a resolution to the measurement problem
and a characterization of when events occur. The resulting
construction is completely formulated in terms of quantum mechanical
objects, without requiring the observation of any classical preferred quantity.
More work is needed in order to fill some gaps related with the proofs of
undecidability in more general contexts and the inclusion of 
interactions between the system and the clock.

The formulation is naturally geared towards dealing with generally 
covariant theories like quantum general relativity. It may also have
implications for how the quantum to classical transition in cosmological
perturbations in the inflationary period take place.

\section{Acknowledgments}

We wish to thank Mario Castagnino, Edgardo Garc\'{\i}a Alvarez,
Luc\'{\i}a Lewowicz, Olimpia Lombardi and Daniel Sudarsky for
discussions and to the anonymous referees for comments.  This work was
supported in part by grant NSF-PHY-0650715, funds of the Hearne
Institute for Theoretical Physics, FQXi, CCT-LSU, Pedeciba and ANII
PDT63/076. This publication was made possible through the support of a
grant from the John Templeton Foundation. The opinions expressed in
this publication are those of the author(s) and do not necessarily
reflect the views of the John Templeton Foundation.

\end{document}